\RequirePackage{lineno}
\documentclass[prl,aps,twocolumn, showpacs,amssymb]{revtex4-1}
\usepackage{t1enc}
\usepackage[latin1]{inputenc}
\usepackage[english]{babel}
\usepackage{graphics}
\usepackage{graphicx}
\usepackage{epsfig}
\usepackage{amssymb}
\usepackage{amsmath}
\usepackage{dcolumn}
\usepackage{bm}
\usepackage{color}
\usepackage{verbatim}  
\usepackage{vector}  
\usepackage{wrapfig}
\usepackage{enumerate}
\usepackage{sidecap}

\usepackage{lineno}

\begin{document}

\renewcommand{\figurename}{Fig.}

\title{Why $T_c$ of (CaFeAs)$_{10}$Pt$_{3.58}$As$_8$  is twice as high as (CaFe$_{0.95}$Pt$_{0.05}$As)$_{10}$Pt$_3$As$_8$}
\author{S.\ Thirupathaiah,$^1$ T.\ St$\ddot{u}$rzer,$^2$ V. B.\ Zabolotnyy,$^1$  D.\ Johrendt,$^2$ B.\ {B{\"u}chner},$^{1,3}$ and S. V.\ Borisenko$^1$}

\affiliation{
$^1$IFW-Dresden, P.O.Box 270116, D-01171 Dresden, Germany\\
$^2$Department Chemie, Ludwig-Maximilians-Universit$\ddot{a}$t M$\ddot{u}$nchen, D-81377 M$\ddot{u}$nchen, Germany\\
$^3$Institut f$\ddot{u}$r Festk$\ddot{o}$rperphysik, Technische Universit$\ddot{a}$t Dresden, D-01171 Dresden, Germany}

\date{\today}

\begin{abstract}
 Recently discovered (CaFe$_{1-x}$Pt$_x$As)$_{10}$Pt$_3$As$_8$ and  (CaFeAs)$_{10}$Pt$_{4-y}$As$_8$ superconductors are very similar materials having the same elemental composition and structurally similar superconducting FeAs slabs.  Yet the maximal critical  temperature achieved by changing Pt concentration is approximately twice higher in the latter. Using angle-resolved photoemission spectroscopy(ARPES) we compare the electronic structure of their optimally doped compounds and find drastic differences. Our results highlight the sensitivity of critical temperature to the details of fermiology and point to the decisive role of band-edge singularities in the mechanism of high-$T_c$ superconductivity. \end{abstract}
\maketitle

A new class of high temperature superconductors~\cite{Kakiya2011,Loehnert2011,Ni2011}, (CaFe$_{1-x}$Pt$_x$As)$_{10}$Pt$_3$As$_8$ (1038) and (CaFePtAs)$_{10}$Pt$_{4-y}$As$_8$ (1048),  in  the family of ironpnictides~\cite{Kamihara2008b,Hsu2008, Rotter2008,Jeevan2008a,Wang2008} has been discovered recently. Although these two compounds are almost similar, both consisting of tetrahedral FeAs planes sandwiched between the planar Pt$_n$As$_8$ (n=3, 4) intermediary layers, differences in their crystal structure and electronic properties have been found. The parent Ca-Pt-Fe-As compound in 1038 phase has a triclinic crystal structure and is semiconducting, whereas the parent 1048 has a tetragonal crystal structure and is metallic~\cite{Ni2011,Loehnert2011} with the band structure calculations pointing to the increased metallicity of the PtAs layers~\cite{Shein2011}. Furthermore, parent 1038 is ordered antiferromagnetically (AFM) below a N$\acute{e}$el temperature 120 K~\cite{Sturzer2013} and unlike in other known ironpnictides the magnetic transition does not lead to any further reduction in the crystal symmetry,  but breaks the tetrahedral symmetry of FeAs layers. 
Superconductivity in 1038 and 1048 systems can be obtained in several ways, either doping FeAs layers directly or via interstitial PtAs layers~\cite{Loehnert2011, Sturzer2012b}. In addition, superconductivity can also be induced by doping electrons through a La substitution at the Ca site~\cite{Sturzer2012b}. By changing only Pt content, up to date a maximum $T_c$ of 38 K is obtained in (CaFeAs)$_{10}$Pt$_{4-y}$As$_8$ superconductor~\cite{Kakiya2011}, and a maximum of 15 K is observed in (Ca$_{1-x}$Pt$_x$FeAs)$_{10}$Pt$_3$As$_8$~\cite{Sturzer2012b}.  
One of the proposals to explain this difference was based on an analogy with the cuprates and attributes high $T_c$s to the increased interlayer hopping induced by high density of states at the Fermi level ($E_F$) from PtAs layers~\cite{Ni2011}. While the recent ARPES data on 1038 suggesting that the interlayer hopping in these materials is weak thus supporting the conjecture~\cite{Neupane2012}, the data on 1048 are still absent. It is therefore interesting to study the electronic structure of both optimally Pt doped materials under the same experimental conditions.

In this Letter we report on the electronic structure and Fermi surface topology of both superconductors by means of high-resolution ARPES. We find that while the electronic structures have the same components as majority of the iron-based superconductors (IBS), i.e., hole pockets near the Brillouin zone center and electron pockets near the zone corner, there are pronounced differences. In particular,  we observe three band-edge singularities located in the immediate vicinity of $E_F$ in 1048 ($T_c$=35 K), where only one is present in the case of 1038 ($T_c$=15 K). We also discuss the possible implications of these findings.

\begin{figure*}[htbp]
	\centering
		\includegraphics[width=0.9\textwidth]{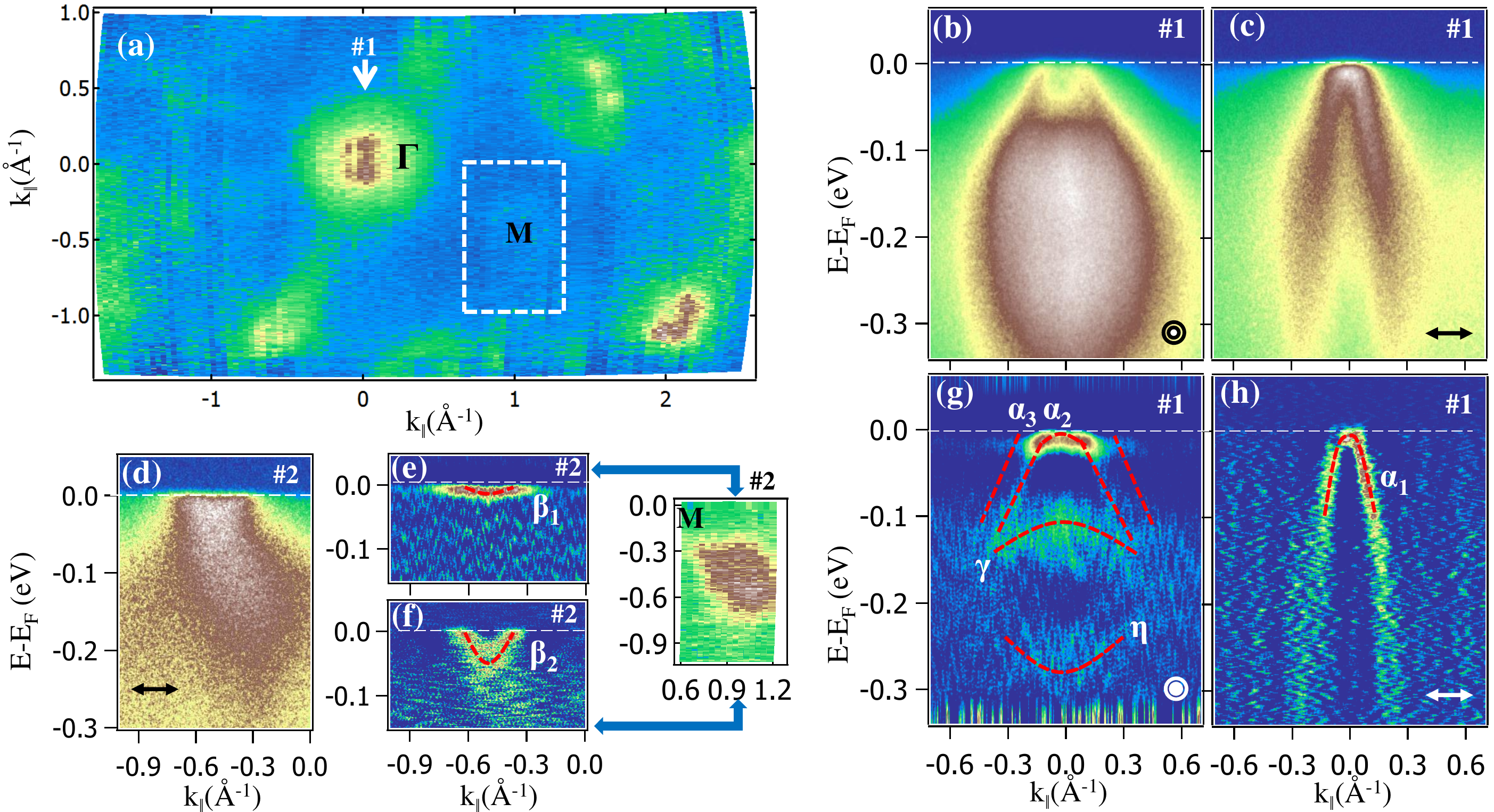}
	\caption{(Color online) ARPES data taken on 1048-35K. (a) shows the FS map measured using $p$-polarized light with an excitation energy of 80 eV. (b) and (c) show  energy distribution map (EDM) cuts taken at $\Gamma$ along the direction shown by arrow on the FS map, measured using $p$- and $s$-polarized lights, respectively. (d) shows EDM cut taken at $M$ along the direction shown by blue arrow on the FS(inset) map, measured using $s$-polarized light. (e) and  (f) are the second derivatives of EDM shown in (d) with respect to momentum and energy, respectively.  (g) and (h) are second derivatives of EDMs shown in (b) and (c), respectively. In the figure double circles and double sided arrow represent $p$- and $s$-polarized lights, respectively.}
\label{1}
\end{figure*}

ARPES measurements were carried out at the UE-112 beam-line equipped with 1$^3$-ARPES end station located in BESSY II (Helmholtz zentrum Berlin) synchrotron radiation center~\cite{Borisenko2012a, *Borisenko2012b}. Photon energies for the measurements were varied between 20 and 80 eV. The energy resolution was set between 5 and 10 meV depending on the excitation energy. Data were recorded at a chamber vacuum of the order of 9 $\times$ 10$^{-11}$ mbar and the sample  temperature was kept at  1 K during the measurements. We employed various photon polarizations in order to probe the symmetry of the electronic bands. The preparation of single crystals, (CaFe$_{0.95}$Pt$_{0.05}$As)$_{10}$Pt$_3$As$_8$ and (CaFeAs)$_{10}$Pt$_{3.58}$ As$_8$, and their elemental analysis are reported elsewhere~\cite{Loehnert2011, Sturzer2013}. The former compound shows superconductivity at a transition temperature $T_c$=15 K, and the latter shows at $T_c$=35 K. From now we refer the former compound by 1038-15K and the latter by 1048-35K.

Figure~\ref{1}(a) shows Fermi surface (FS) map of superconducting 1048-35K compound measured using $p$-polarized light with an excitation energy $h\nu$=80 eV. The map was extracted over an integration window larger than the size of superconducting gap so that its influence on the intensity distribution is negligible. In order to identify all features of the electronic structure, which sometimes are hidden because of the strong influence of the matrix element effects, one needs to record the map in a big portion of the $k$-space and use the light of various polarizations and energies. We therefore have measured a part of the map using $s$-polarized light and detected the "missing" spectral weight (white dashed contour). In the map we could thus observe the signatures of the hole pockets at the zone centers [near (0,0), (1.5, 0.5), (2,-1) and partially near (0.5, -1.5), (-1.5, -0.5)] and electron pockets at the zone corners [ near (1, -0.5), (-0.5, -1), (-1, 0.5) and (0.5, 1)], as is typically the case for all IBSs.

The spectral intensity near the zone center is formed by holelike features as shown in the energy distribution maps (EDM) [see Figs.~\ref{1}(b) and ~\ref{1}(c)].  From the figures we can clearly see that only one of the three features, $\alpha_3$, crosses $E_F$ and thus alone contributes to the Fermi surface. We calculated a  Fermi vector $k_F \approx $  0.29 \AA$^{-1}$ ~and Fermi velocity $v_F \approx$ 0.4 eV-\AA ~for the band $\alpha_3$. The two other bands, $\alpha_1$ and $\alpha_2$, only approach $E_F$  but do not cross it. This observation is similar to the case of LiFeAs at particular $k_z$ values (for instance see Fig.~2(d) in Ref.~\onlinecite{Borisenko2010}). With the help of polarization dependent measurements, near $\Gamma$ we could assign Fe 3d$_{yz , xz, xy}$ orbital characters to $\alpha_1$, $\alpha_2$, and $\alpha_3$ bands, respectively~\cite{Thirupathaiah2011,Fink2009}. In Fig.~\ref{1}(g) one can observe that the top of $\alpha_2$ forms a band-edge singularity at $E_F$ and this, as we will see later, is remarkably different from 1038-15K compound (see Fig.~3). We further notice two more bands $\gamma$ and $\eta$ at the zone center as shown in Fig.~\ref{1}(b). The former is located at around 105 meV below $E_F$ and disperses towards higher binding energy, while the latter disperses towards lower binding energy and its bottom lies at around 280 meV from $E_F$. The $\gamma$ band is most likely resulted from a hybridization between Pt $5d$ and As $4p$ in the Pt$_n$As$_8$ layers~\cite{Shein2011}, while the $\eta$ band originates from Fe 3d$_{z^2}$ states~\cite{Thirupathaiah2011}.

From the FS maps we can see that the electron pockets are elongated in the $\Gamma-M$ direction, indicating that these pockets are shallow. Indeed, this is further confirmed by the panels ~\ref{1}(d), ~\ref{1}(e), and ~\ref{1}(f).  In Figs.~\ref{1}(e) and ~\ref{1}(f) we can clearly see the presence of shallow ($\beta_1$) and deep ($\beta_2$) electron pockets near the zone corner, where one of them ($\beta_1$) forms a band-edge near $E_F$. Thus, another singularity is located at the zone corner close to $E_F$ in the case of 1048-35K. The presence of hole pockets at the zone center and electron pockets at the zone corner is in consistent with the electron structure of other ironpnictides~\cite{Lu2008a, Ding2008a, Xia2009a}. Apart from the above mentioned features we do not observe any spectral intensity near $E_F$ that could ratify the Pt-related states, which is in good agreement with the reported DFT calculations ~\cite{Loehnert2011}.

	\begin{figure}[t]
		\centering
			\includegraphics[width=0.48\textwidth]{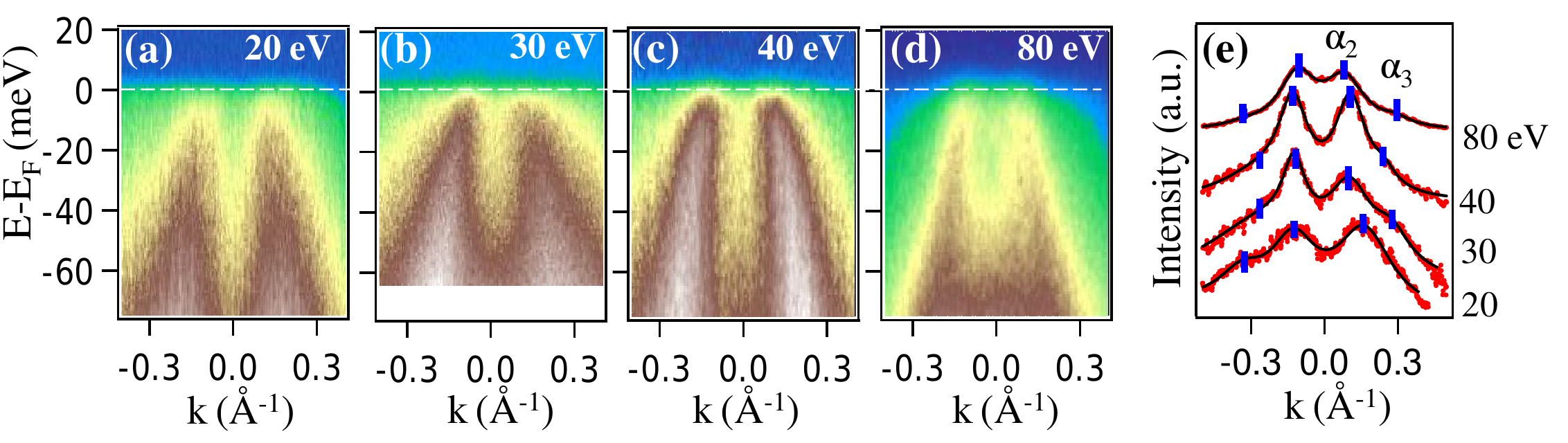}
		\caption{(Color online) ARPES data taken on 1048-35K. (a-d) show  photon energy dependent EDMs measured at the zone center. (e) shows  momentum distribution curves taken over an integration range of 10 meV with respect to the Fermi level, derived from EDMs shown in (a-d). Vertical blue lines in the figure (e) represent peak positions of the bands $\alpha_2$ and $\alpha_3$.}
		\label{2}
	\end{figure}

In Fig.~\ref{2} we show photon energy dependent measurements performed to reveal band dispersion in the $k_z$  direction. Figs.~\ref{2}(a-d) show EDMs measured at the zone center using photon energies 20, 30, 40, and 80 eV and corresponding $k_z$ values are given by 2.13 $\pi/c$, 2.53 $\pi/c$, 2.87 $\pi/c$, and 3.94 $\pi/c$ ($c$=10.46 \AA),  respectively. From  these $k_z$ values it is clear that the photon energies 20 and 80 eV detect the bands at $Z$-point, while 30 eV detects them at $\Gamma$-point. In Fig.~\ref{2}(e) we show momentum distribution curves (MDCs) taken from the EDMs shown in  Figs.~\ref{2}(a-d) over an integration range of 10 meV with respect to $E_F$. As obtained MDCs are fitted using a function of four Lorentzians. From the fits we extracted peak position of the bands $\alpha_2$ and $\alpha_3$, marked by vertical blue lines as shown in Fig.~\ref{2}(e). We further notice peak positions of $\alpha_2$ and  $\alpha_3$ at high the symmetry points $\Gamma$ and $Z$ are almost equivalent within the experimental errors, which suggests a quasi-2D hole dispersion along the $\Gamma-Z$ direction for these compounds. If the interlayer coupling between FeAs layers was enhanced by the intermediary PtAs planes, as proposed in Ref.~\onlinecite{Ni2011}, we should observe a strong dispersion of the bands along the $\Gamma-Z$ direction~\cite{Thirupathaiah2011}. On the contrary, we observe typically weak (but finite) $k_z$ dispersion for the hole pockets. We further observe a low spectral weight for $\alpha_3$. This could be due to the dominant in-plane $xy$ orbital contribution to $\alpha_3$,  as they have low scattering cross-sections in the photoemission process~\cite{Fink2009,Thirupathaiah2012}.
	
\begin{figure}[t]
		\centering
			\includegraphics[width=0.48\textwidth]{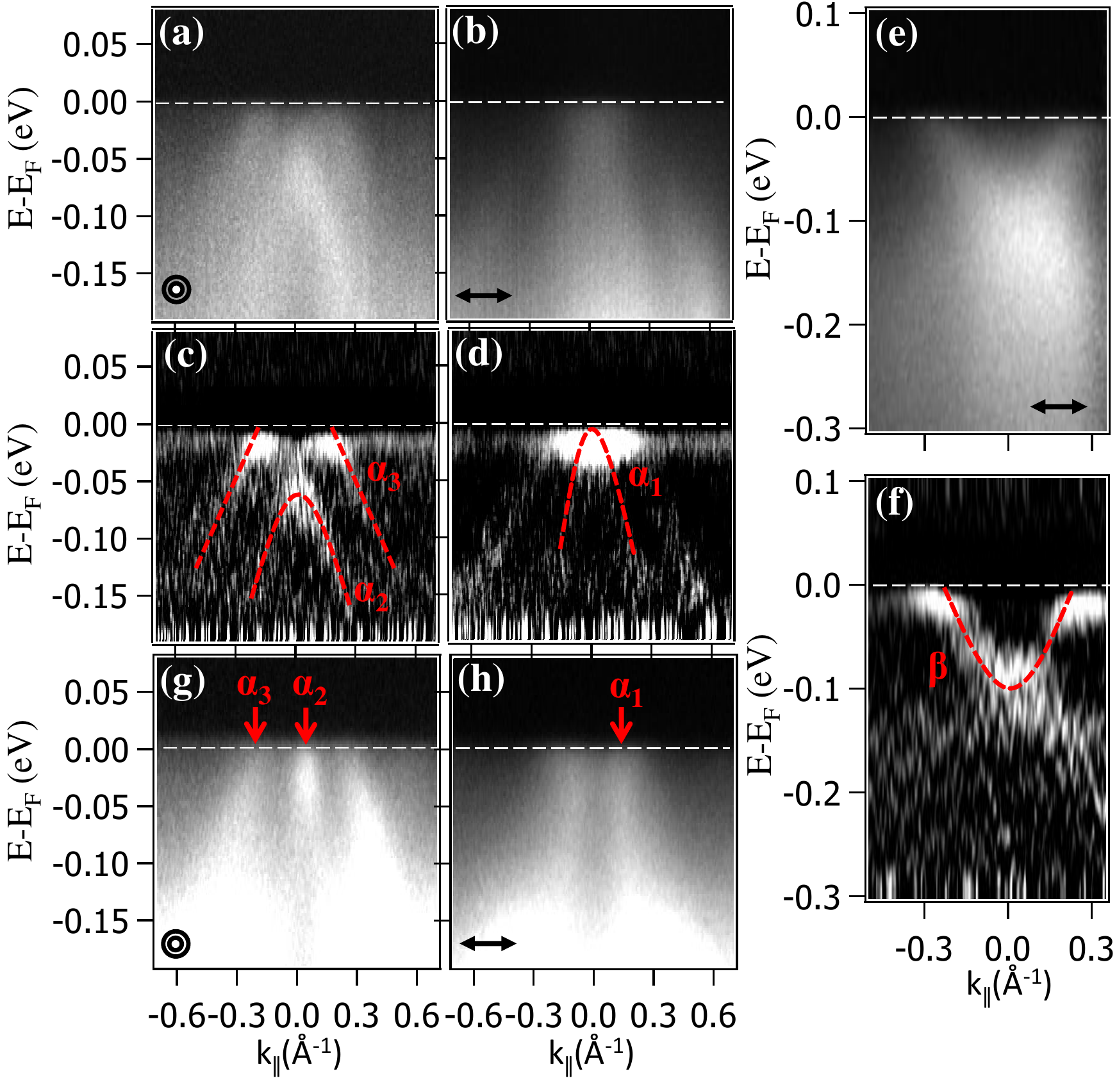}
		\caption{(Color online) ARPES data taken on 1038-15K. (a) and (b) show  EDMs taken at the zone center measured using an excitation energy of 70 eV with $p$- and $s$-polarized lights, respectively. (c) and (d) are second derivatives of  EDMs shown in (a) and (b), respectively. (e) is  EDM taken at the zone corner and (f) is second derivative (e). (g) and (h) are EDMs measured with the same photon energy and polarization of (a) and (b), respectively but from a different polar angle which corresponds to a different $k_z$. In the figure double circles and double sided arrow represent $p$- and $s$-polarized lights, respectively.}
		\label{3}
	\end{figure}
		
Next, in Fig.~\ref{3} we show the ARPES data taken on the superconducting 1038-15K compound. All the data were recorded with an excitation energy of 70 eV. We observe three holelike band dispersions $\alpha_1$, $\alpha_2$,  and $\alpha_3$ at the zone center [see Figs.~\ref{3}(a) and ~\ref{3}(b)] and an electronlike band dispersion $\beta$ at the zone corner [see Fig.~\ref{3}(c)]. In Figs.~\ref{3}(a) and ~\ref{3}(c) we observe top of the band $\alpha_2$ is approximately  75 meV below $E_F$, while $\alpha_1$ disperses to $E_F$ and touches it. Therefore, only $\alpha_3$ crosses $E_F$ in this 70 eV data ($k_z$ = 7.26 $\pi/c$) near zone center. We calculated a Fermi vector $k_F$=0.18 \AA$^{-1}$ ~and Fermi velocity $v_F$=0.45 eV-\AA ~for $\alpha_3$.  In order to confirm the presence of all three bands, we also show the data [Figs.~\ref{3}(g) and ~\ref{3}(h)] on hole pockets from a different polar angle measured with the same photon energy which corresponds to a different $k_z$. Our 1038-15K sample appears to be more electron doped than the one presented in Ref.~\onlinecite{Neupane2012}, where already two holelike bands crossed the Fermi level. It is clearly seen in Figs.~\ref{3}(e) and \ref{3}(f) that both electron pockets (though not clearly resolved) have their band bottoms much farther away from $E_F$ than in 1048-35K.
				
\begin{figure}
		\centering
			\includegraphics[width=0.48\textwidth]{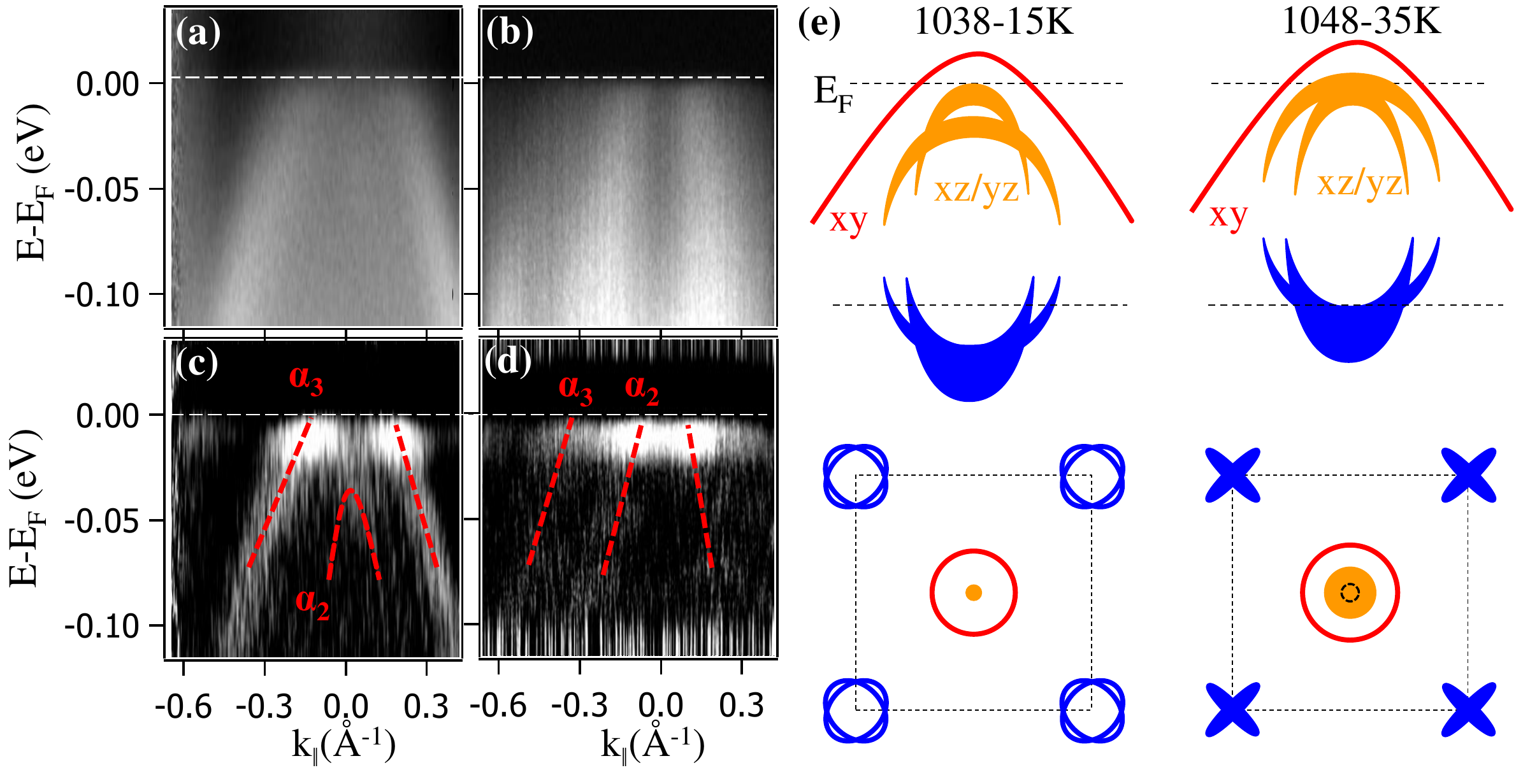}
		\caption{(Color online) (a) and (b) are EDMs measured using $p$-polarized light with an excitation energy of 20 eV from 1038-15K and 1048-35K, respectively. (c) and (d) are second derivatives of (a) and (b), respectively. (e) shows schematic representation of the experimentally determined electronic structure near $\Gamma$ for both compounds.}
		\label{4}
	\end{figure}

We now switch to comparison of the electronic structure between higher- and lower-$T_c$ compounds in order to identify the basic elements for enhancement in the critical temperature. In Fig.~\ref{4} we show EDMs taken at the zone center measured using $p$-polarized light with an excitation energy of 20 eV. From previous discussion it is clear that this photon energy extracts the states near the $Z$-point.  
 We can see that the band $\alpha_3$ crosses $E_F$ in both cases, while supporting a slightly larger Fermi surface in 1048-35K. On the other hand in 1038-15K the  top of the band $\alpha_2$ is much deeper ($\approx$40 meV from $E_F$), therefore,  hardly playing any role in transport or superconductivity. In contrast,  $\alpha_2$ in 1048-35K approaches $E_F$ and its top virtually coincides with it for a range of momenta.  We calculated $v_F$=0.33 eV-\AA ~and $k_F$=0.15 \AA$^{-1}$ ~for $\alpha_3$ in 1038-15K. Similarly, we found $v_F$=0.3 eV-\AA ~and $k_F$=0.25 \AA$^{-1}$ ~for $\alpha_3$ in 1048-35K. Constant Fermi velocity of the band $\alpha_3$ suggests similar in-plane interactions in both compounds. By shifting the Fermi level in Fig.~\ref{4}(a) up to 55 meV towards higher binding energy we can reproduce Fig.~\ref{4}(b). At a glance one can think of rigid-band scenario in these compounds.  However, the presence of $\alpha_1$ close to $E_F$ in both 1038-15K and 1048-35K [see Figs.~\ref{1}(c) and ~\ref{3}(b)] excludes such a behaviour.  Currently we cannot offer a reasonable explanation for this effect, but we do not exclude a possible influence of the triclinic crystal structure of the 1038 compounds which can lift the degeneracy of the $xz, yz$ states at the $\Gamma$-point, which otherwise can be lifted only by inclusion of the spin-orbit interaction~\cite{Ma2010}.

Another clear difference between these two compounds is the total charge in the FeAs slabs. Since the $\alpha_3$ Fermi sheet is slightly larger and electron pockets are significantly shallower in 1048-35K, it appears that this compound is more hole-doped compared to 1038-15K. Although (CaFeAs)$_{10}$Pt$_{3.58}$ As$_8$ (1048-35K) has higher net Pt, in CaFe$_{0.095}$Pt$_0.05$As)$_{10}$Pt$_3$As$_8$ (1038-15K) it is substituted directly into the FeAs slabs which affects the electronic structure more than in the former where Pt is substituted into the intermediary PtAs layers. Note that the critical temperatures of transition-metal doped compounds are generally much lower than those of charge doped or pure FeAs-layers.

Comparing our experimental observations with the published band-structure calculations of these compounds~\cite{Shein2011}, we notice a partial agreement. The experimental hole dispersions at the zone center qualitatively agree with the predicted quasi-2D hole pockets along the $k_z$ direction. On the other hand, the predicted electronlike band at the zone corner which has a sharp dispersion along $k_z$ is not observed experimentally in our measurements. Absence of such a band reduces the chance of high degree of interlayer coupling that is argued as a reasonable explanation for the pairing mechanism in these compounds~\cite{Ni2011}.
	
 In Fig.~\ref{4}(e) we schematically show the experimentally determined electronic structure for both compounds. Here, the key difference is obviously the number of band-edges near $E_F$ in higher $T_c$ superconductor (1048-35K) compared to the lower one (1038-15K). Such singularities, provided that the band-edges are not parabolic, can induce high density of states near $E_F$ and thus enhance $T_c$.  These band-edge singularities do exist in every IBS with a significant critical temperature~\cite{Zabolotnyy2009a,Borisenko2010,Liu2010b,Borisenko2012}, which appear to be eminent for acquiring superconductivity in ironpnictides. In this study, the presence of two additional band-edges near the Fermi level in 1048-35K is a plausible explanation for more robust pairing compared to 1038-15K where we observe only one band-edge singularity. Moreover, in 1038-15K the nesting conditions are not perfect due to different sizes of the hole and electron pockets, and also the intraorbital interactions are suppressed since the "active" $xz,yz$ states are far below $E_F$. Whereas in 1048-35K the $xz, yz$ states are close to $E_F$ at both zone center and corner, showing the necessity of these states to present as many as near $E_F$ for high $T_c$ superconductivity in ironpnictides.


In conclusion, we have studied the electronic structure of newly discovered Ca-Pt-Fe-As-type ironpnictide superconductors in both 1038 and 1048 phases. We observed, typical for IBS, holelike bands at the zone center and electronlike bands at the zone corner for both compounds. While the degree of interlayer coupling was found similar in both materials, the pronounced difference in the low-energy electronic structure itself could offer a very plausible explanation for the enhanced critical temperature in 1048-35K compared to 1038-15K. Three band-edge singularities present in the immediate vicinity of the Fermi level in  the system that has higher $T_c$ (1048-35K) where only one is realized for lower $T_c$ (1038-15K) compound. Our experimental findings underline the importance of fine tuning of the electronic structure within the FeAs layers by interstitial atoms, as well as the role played by the band-edge singularities in the mechanism of high $T_c$ superconductivity in iron-based superconductors. Our results provide no evidence for the strong interlayer coupling among the FeAs layers.
	
\bibliographystyle{apsrev}
\bibliography{Pt}
\end{document}